\documentclass{iaus}
\usepackage{graphicx}

\def\arcsec{\hbox{$^{\prime\prime}$}}

\def\msun{M$_{\odot}$}
\def\msunpyr{M$_{\odot}$ yr$^{-1}$}

\def\d4000{$D_{\rm 4000}$}
\def\halpha{\ifmmode {\rm H{\alpha}} \else $\rm H{\alpha}$\fi}
\def\hbeta{\ifmmode {\rm H{\beta}} \else $\rm H{\beta}$\fi}
\def\hda{\ifmmode {\rm H{\delta}_{\rm A}} \else $\rm H{\delta}_{\rm A}$\fi}
\def\hd{\ifmmode {\rm H{\delta}} \else $\rm H{\delta}$\fi}
\def\oii{[O\,{\sc ii}]$\lambda$3727}

\def\oiiib{[O\,{\sc iii}]$\lambda$5007}

\def\niib{[N\,{\sc ii}]$\lambda$6584}

\def\sii{[S\,{\sc ii}]$\lambda\lambda$6717,6731}
\def\nn2{$N2$}
\def\rr23{$R_{\rm 23}$}
\def\oo32{$O_{\rm 32}$}
\def\oo2ne3{$O_{\rm 2Ne3}$}

\title[Mass Assembly Survey with SINFONI in VVDS] 
{Probing the Mass Assembly and Chemical Evolution of High-z Galaxies with MASSIV}

\author[T. Contini et al.]   
{T. Contini$^{1,2}$, 
B. Epinat$^{1,2}$, 
J. Queyrel$^{1,2}$, 
D. Vergani$^3$, 
L. Tasca$^4$, 
P. Amram$^4$, 
B. Garilli$^5$, 
M. Kissler-Patig$^6$, 
O. Le F\`evre$^4$, 
J. Moultaka$^{1,2}$, 
L. Paioro$^5$, 
L. Tresse$^4$, 
F. Bournaud$^7$, 
 \and E. Perez-Montero$^8$
}

\affiliation{
$^1$IRAP - CNRS, 14, avenue Edouard Belin, F-31400 Toulouse, France \\
$^2$IRAP-Universit\'e de Toulouse-UPS-OMP, Toulouse, France; email: {\tt contini@ast.obs-mip.fr}  \\
$^3$ INAF - Osservatorio Astronomico di Bologna, via Ranzani 1, 40127 Bologna, Italy \\
$^4$ LAM, UMR6110, CNRS-Universit\'e de Provence Aix-Marseille I, Marseille, France \\
$^5$ IASF-INAF, via Bassini 15, 20133 Milano, Italy \\
$^6$ European Southern Observatory, Karl-Schwarzschild-Strasse 2, 85748 Garching, Germany \\
$^7$ CEA, IRFU, SAp, 91191 Gif-sur-Yvette, France \\
$^8$ Instituto de Astrofisica de Andalucia, CSIC, Apartado de Correos 3004, 18080Granada, Spain \\
}

\pubyear{2011}
\volume{272}  
\pagerange{119--126}
\jname{Tracing the Ancestry of Galaxies: On the Land of Our Ancestors}
\editors{Carignan, C., Freeman, K. C.  \&  Combes, F., eds.}
\begin{document}

\maketitle

\begin{abstract}
Understanding the different mechanisms of galaxy assembly at various cosmic epochs is a key issue for galaxy evolution 
and formation models. We present MASSIV (Mass Assembly Survey with SINFONI in VVDS) in this context, an on-going 
survey with VLT/SINFONI aiming to probe the kinematics and chemical abundances of a unique sample of 84 star-forming 
galaxies selected in the redshift range $z\sim 1-2$. This large sample, spanning a wide range of stellar masses, is 
unique at these high redshifts and statistically representative of the overall galaxy population. In this paper, we give an 
overview of the MASSIV survey and then focus on the spatially-resolved chemical properties of high-z galaxies and their 
implication on the process of galaxy assembly. 
\end{abstract}

\firstsection 
\section{Introduction}
The details of galaxies assembly and evolution processes still remain 
relatively  unknown. Much of our current knowledge of the high-redshift galaxy 
populations still relies on the integrated spectra and multi-wavelength photometry 
acquired through deep and wide surveys (eg. VVDS and COSMOS). 
More detailed constraints are however needed to understand the main physical processes 
(angular momentum exchange, dissipation and cooling, feedback from star formation or AGN, etc) 
involved in the formation and evolution of galaxies. Such constraints are
indeed crucial inputs for theories and simulations of galaxy formation and evolution.
Spatially-resolved investigations of individual galaxies at
early stages of their evolution is thus crucial to disentangle the different processes 
of galaxy mass assembly. Thanks to the advent of sensitive near-infrared (NIR)  integral field spectrographs (eg. SINFONI, OSIRIS) 
mounted on 8-10m class telescopes such studies have been recently made possible (see Epinat et al. in these proceedings for a review), 
allowing to probe the complex 
kinematics and morphologies of high-redshift galaxies, and enabling the mapping 
of the distribution of star formation and physical properties
such as chemical abundances. 
These powerful instruments indeed provide access, 
for $z\sim 1-4$ galaxies, to the well-calibrated spectral
diagnostics of the physical properties from rest-frame optical
emission lines such as \halpha, \hbeta, \niib, \sii, \oiiib, and \oii.
Using the NIR integral field spectrograph SINFONI at ESO/VLT, we are conducting a major survey 
of spatially-resolved studies of high-redshift galaxy populations: the
Mass Assembly Survey with SINFONI in VVDS, or "MASSIV".  
With the detailed information provided by SINFONI on individual galaxies, 
the key science goals of the MASSIV survey are to investigate in detail: 
(1) the nature of the dynamical support (rotation vs. dispersion) of high-z 
galaxies, (2) the respective role of mergers (minor and/or major) and gas 
accretion in galaxy mass assembly, (3) and the process of gas exchange 
(inflows/outflows) with the intergalactic medium through the derivation of 
metallicity gradients.
The MASSIV sample includes 84 star-forming galaxies drawn from 
the VIMOS VLT Deep Survey (VVDS) in the redshift range $0.9 < z < 1.8$. 
So far, we have collected and reduced observations of 50 galaxies. 
In this paper, we present the MASSIV sample together with the first results coming out 
from the analysis of the "first epoch" 50 MASSIV galaxies. Throughout this paper, 
we assume a standard $\Lambda$-CDM cosmology, i.e. $h=0.7$,
$\Omega_{\mathrm{m}}=0.3$ and $\Omega_{\Lambda}=0.7$. 
For this cosmology, 1\arcsec\ corresponds to $\sim 8$ kpc at $z\sim 1-2$. 

\section{The MASSIV sample: selection criteria and global properties}

\begin{figure}[bt]
\begin{center}
 \includegraphics[width=8cm]{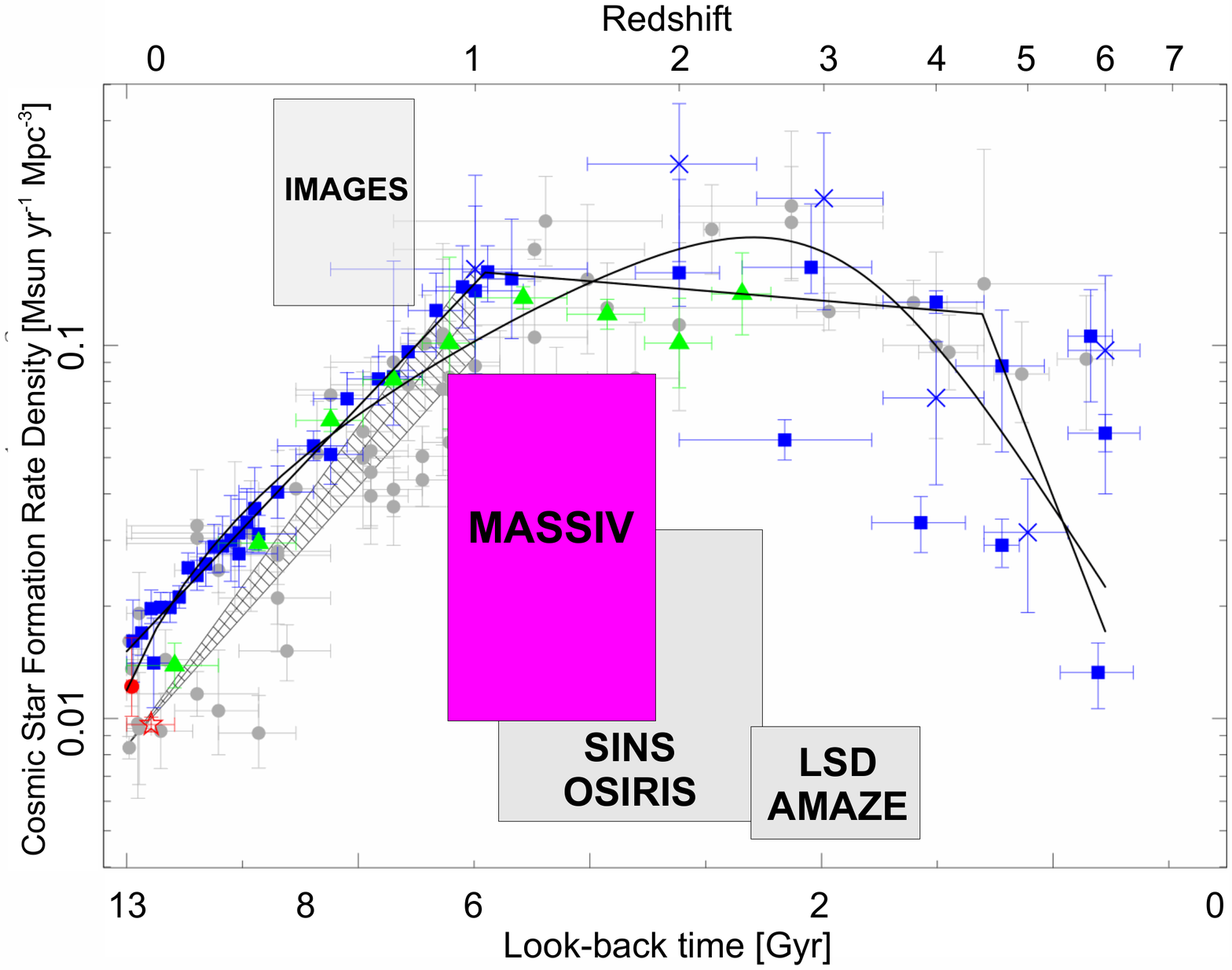}
 \includegraphics[width=5.4cm]{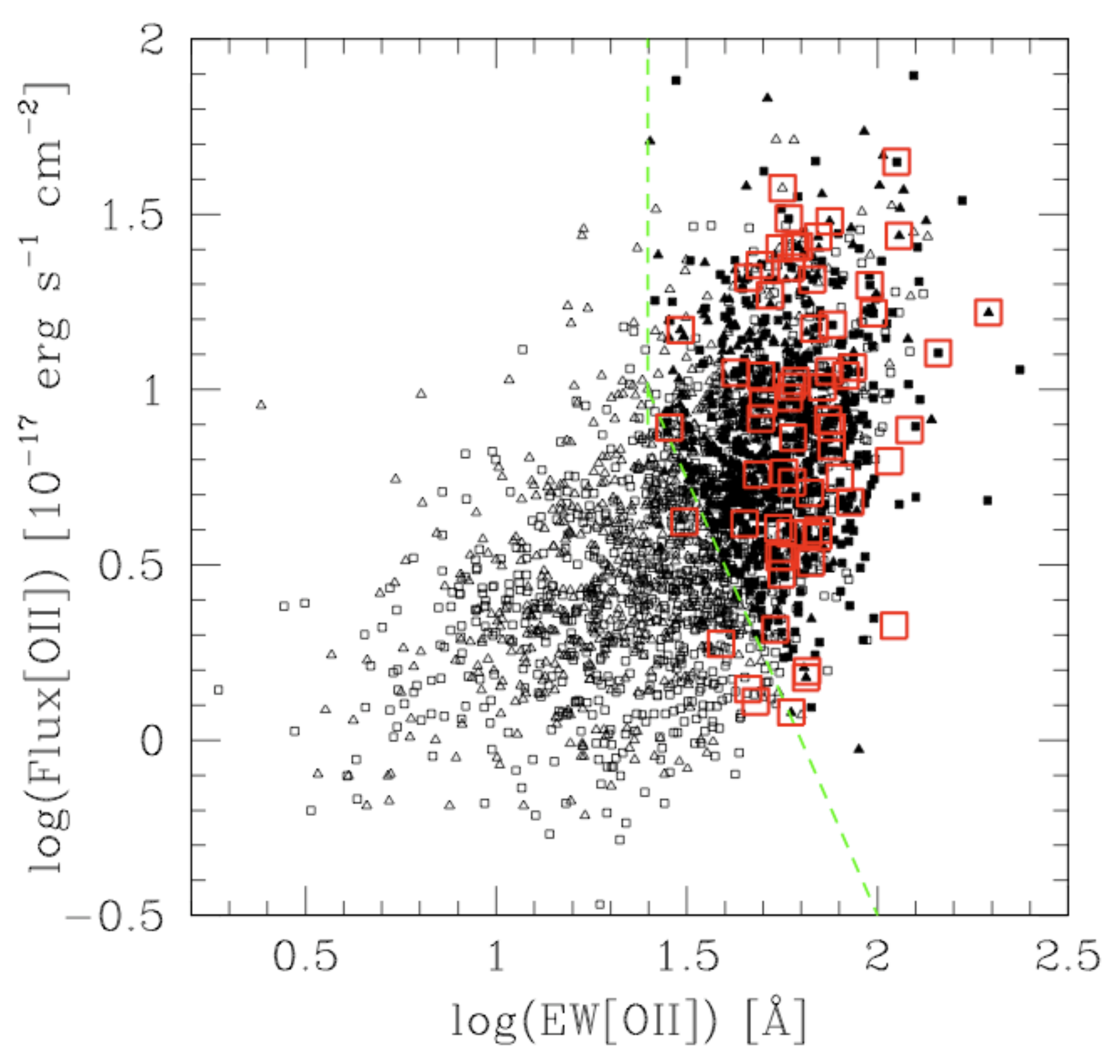}
      \caption{{\it Left}. Evolution of the cosmic star formation rate density as a function of look-back time and redshift (adapted from Hopkins 2006). The redshift range of MASSIV ($0.9 < z < 1.8$, magenta box) is compared with other major IFU surveys of distant .galaxies: IMAGES ($z \sim 0.4-0.8$), SINS/OSIRIS ($z \sim 1.4-2.6$), and LSD/AMAZE ($z \sim 2.6-3.8$). The relative height of each boxes is proportional to the samples size. {\it Right}. Selection of star-forming galaxies with secure redshift 
and measured \oii\ emission line in the VVDS fields for SINFONI follow-up observations. The green dashed line indicates the selection box of MASSIV targets (filled symbols). The 64 galaxies selected for the MASSIV survey based on their \oii\ emission-line strength are indicated as red squares.                                }
         \label{zhist_complitt}
\end{center}
\end{figure}

We have used the VVDS sample to select galaxies across the peak of star formation activity around $z\sim 1.5$. 
VVDS offers the advantage of combining a robust selection function and secure spectroscopic redshifts. For the MASSIV survey, we have defined a sample of 84 VVDS star-forming galaxies at $0.9 < z < 1.8$ suitable for SINFONI observations. Three selection crtiteria have been applied successively. First, the MASSIV targets were selected to be star-forming galaxies. In most of the cases, the selection was based on the measured intensity of \oii\ emission line in the VIMOS spectrum (see Figure~\ref{zhist_complitt}) or, for a few cases where the \oii\ emission line is out of the VIMOS spectral range, on their observed photometric $UBVRIK$ spectral energy distribution (SED) and/or UV rest-frame spectrum which is typical of star-forming galaxies. The star formation criteria ensure that the brightest rest-frame optical emission lines, mainly \halpha\ and \niib\ used to probe kinematics and chemical abundances, will be observed with SINFONI in the NIR $J$ and $H$ bands. Among these star-forming galaxy candidates, we have further restricted the sample taking into account one important observational constraint: the observed wavelength of H$\alpha$ line had to fall at least 9\AA\ away from strong OH night-sky lines to avoid heavy contamination of the galaxy spectrum by sky subtraction residuals.
Finally, a fraction of MASSIV galaxies have been selected to be observed at higher spatial resolution with the adaptive optics (AO) system of SINFONI assisted with the Laser Guide Star facility. In these cases, a bright star ($R < 18$ mag) close enough to the target ($d < 60\arcsec$) is needed for the 
zero-order tip-tilt corrections.  Most of the MASSIV galaxies were observed in seeing-limited mode with a median seeing of $\sim 0.65$\arcsec. Ten targets only have been acquired with AO, achieving FWHM resolution of $\sim 0.25$\arcsec. The total on-source integration times range on average from 80 to 120 minutes (see Contini et al., in prep. for details). 
MASSIV galaxies are distributed in the redshift range between $z\sim 0.94$ and $1.80$ with a median value $z=1.33$. MASSIV is thus probing a lower redshift range than the SINS, OSIRIS and LSD/AMAZE surveys (see Figure~\ref{zhist_complitt}). However, even if the LSD/AMAZE sample targets galaxies at the highest redshifts  ($z \sim 2.6-3.8$), the MASSIV, OSIRIS and SINS surveys are probing the common redshift range $z \sim 1.3-1.8$. 
Stellar masses and SED-based Star Formation Rates (SFR) for MASSIV galaxies have been derived using standard SED fitting procedures (see Contini et al., in prep. for details). Stellar masses range between $\sim 3 \times 10^{9}$ and $6 \times 10^{11}$ \msun, with a median value of $1.4 \times 10^{10}$ \msun. The stellar mass range probed by MASSIV is rather similar to the one targeted by the SINS, OSIRIS, and LSD/AMAZE surveys, extending from $\sim 10^9$ to $5 \times 10^{11}$ \msun. The median value of the MASSIV sample is intermediate between the median values of the OSIRIS/LSD/AMAZE ($1.1 \times 10^{10}$ \msun) and the SINS ($2.5 \times 10^{10}$ \msun) samples. 
SED-based SFRs of MASSIV galaxies range between $\sim 5$ to $400$ \msunpyr, with a median value of $\sim 31$ \msunpyr. The SFR range probed by MASSIV is rather similar to the one targeted by the SINS, and OSIRIS surveys, extending from $\sim 5$ to $400$ \msunpyr. The median value of the MASSIV sample is however smaller than the median values of the OSIRIS (47 \msunpyr), the SINS (72 \msunpyr), and LSD/AMAZE (100 \msunpyr) samples.

\section{Evidence for positive metallicity  gradients in high-redshift galaxies}

\begin{figure}[bt]
\begin{center}
 \includegraphics[width=6.7cm]{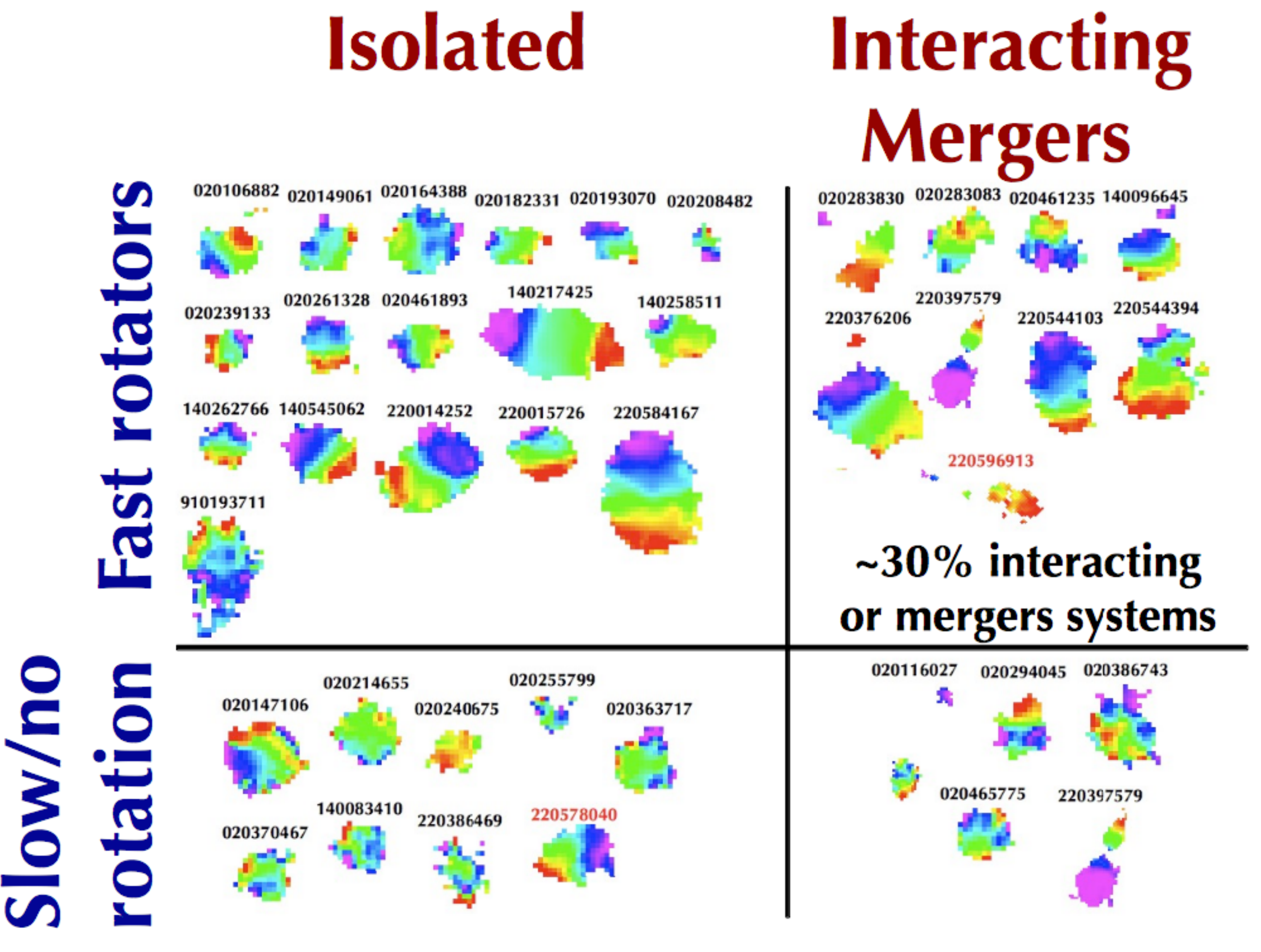}
 \includegraphics[width=6.7cm]{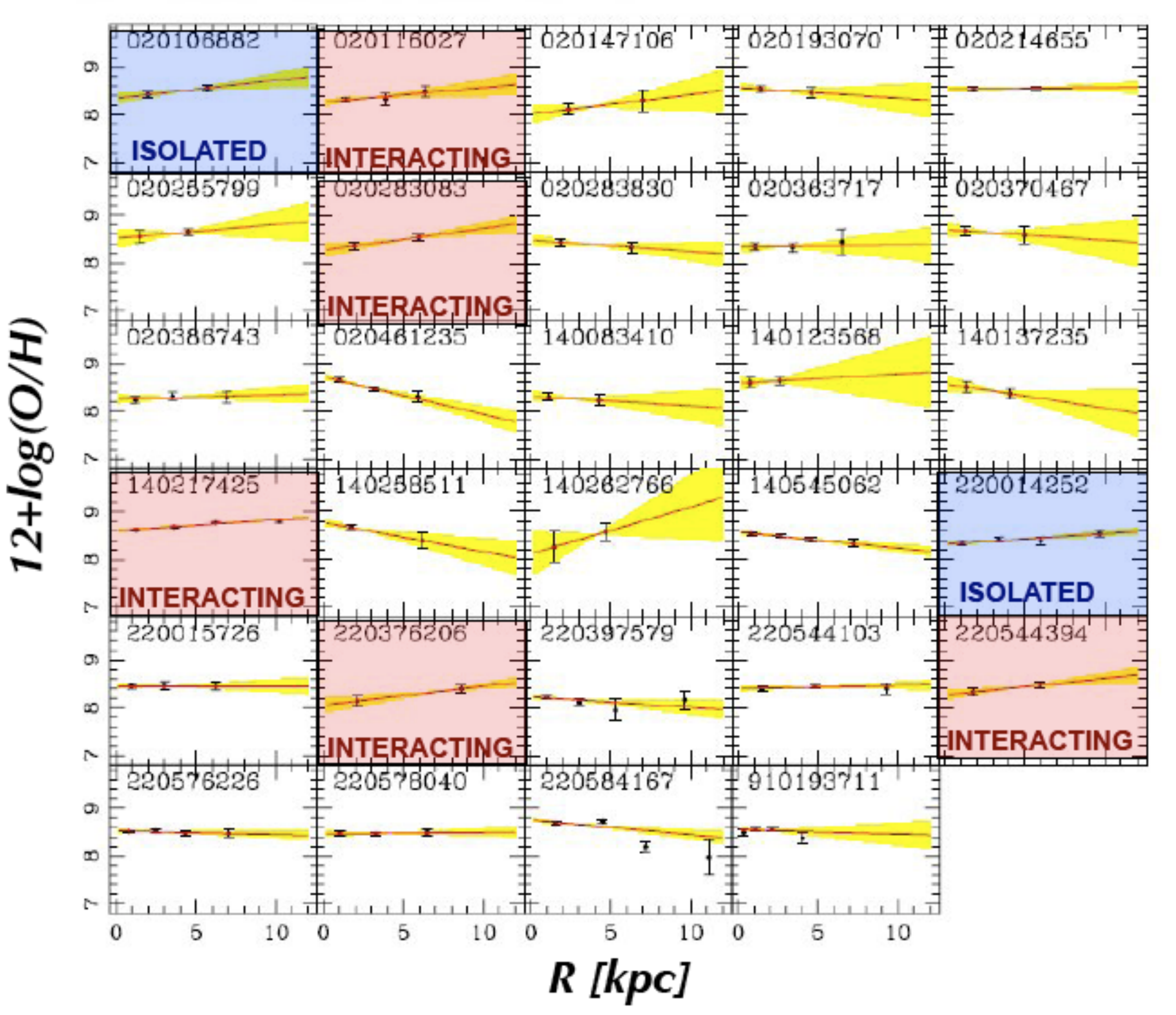}
      \caption{{\it Left}. \halpha-based velocity fields of the "first epoch" MASSIV sample. Galaxies have been classified  in four broad classes depending on their close environment (isolated vs. interacting/merger system) and kinematics (rotating disks vs. slow/no rotation). About one third of MASSIV galaxies show clear signatures of close interactions and/or ongoing merging (see Epinat et al., in prep. for details). {\it Right}. Metallicity gradients for 29 MASSIV galaxies with spatially-resolved metallicities. 
      The red lines are the best fits to the data and the yellow regions represent the 1$\sigma$ errors associated to the gradients. Positive gradients are clearly observed in 7 MASSIV galaxies, among which 5 are interacting systems (see Queyrel et al., in prep.).
                                }
         \label{gradZ}
\end{center}
\end{figure}

A visual classification has been performed on the "first epoch" sample of 50 MASSIV galaxies using mainly CFHT $I$-band images, and velocity field/dispersion maps 
(see Epinat et al., in prep. for details). For each galaxy, a velocity model has been fitted to its velocity map, and a derived velocity dispersion map 
(corrected from beam smearing, see Epinat et al. 2009) is produced. The classification has been performed by eight people independently and then reconcilied to 
produce the final classification with corresponding confidence levels. This broad classification is based on two criteria. The first one concerns the kinematical state of the main 
component and the second one the close environment of the galaxy. We thus defined three types of kinematical classes: disk in regular rotation, perturbed rotation and
quasi non-rotating object (perturbed and rotating), and two types of environment: isolated and interacting/merging objects. The result of this classification is shown in Figure~\ref{gradZ} 
(left). 

At high redshifts the gas-phase metallicity of individual galaxies can only be measured with limited accuracy. Here, we use the \niib/\halpha\  emission-line ratio as an indicator
of oxygen abundance and the calibration obtained by P\'erez-Montero \& Contini (2009). For 29 galaxies of the "first epoch" MASSIV sample, we have been able
to quantify the radial behavior of the gas-phase metallicity (see Fig.~\ref{gradZ}, right). The radial gradients are more often very weak, some being positive and other negative. 
Contrary to the global trend in the local universe, where the gas-phase metallicity of spiral galaxies decreases with galactocentric radius, an important fraction of our galaxies have larger metallicities at larger distance from the center. However, it is not the first time that positive metallicity gradients are found. Recently, Werk et al. (2010) reported a positive gradient in a local galaxy and proposed several scenari to explain their discovery: (i) radial redistribution of the metal-rich gas produced in the nucleus, 
(ii) supernova blowing out metal-rich gas, enriching the IGM, then falling onto the outer parts of the disk, 
(iii) result of a past interaction. If we stick strictly to the numbers, our sample counts 57\% of positive gradients, of course some of them being very weak. 
Taking into account the $1 \sigma$ error on each slope (as shown in yellow in Figure~\ref{gradZ}), we count 7 galaxies with a secure positive gradient. 
Among these galaxies, 5 are classified as interacting systems, while the 2 other are isolated. We thus conclude that the majority of the galaxies for
which we detected a secure positive metallicity gradient are interacting. This result balances the recent interpretation by Cresci et al. (2010) of positive metallicity 
gradients in three $z\sim 3$ galaxies as evidence for cold gas accretion as the main mechanism of galaxy mass assembly at high redshifts. 

\begin{acknowledgements}
This work is supported by the french ANR grant ANR-07-JCJC-0009, the CNRS-INSU and its Programme National Cosmologie-Galaxies.
\end{acknowledgements}


\vspace{-0.2cm}
\begin{thebibliography}{}

\bibitem[Cresci et al.(2010)]{2010Natur.467..811C} Cresci, G., et al.\ 2010, Nature, 467, 811 
\bibitem[Epinat et al.(2009)]{2009A&A...504..789E} Epinat, B., et al.\ 2009, A\&A, 504, 789 
\bibitem[Hopkins(2006)]{} Hopkins, A.~M..\ 2006, astro-ph/0611283
\bibitem[P{\'e}rez-Montero \& Contini(2009)]{2009MNRAS.398..949P} P{\'e}rez-Montero, E., \& Contini, T.\ 2009, MNRAS, 398, 949 
\bibitem[Werk et al.(2010)]{2010ApJ...715..656W} Werk, J.~K., et al.\ 2010, ApJ, 715, 656 


\end{thebibliography}
\end{document}